**Влияние толщины базы на эффективность фотопреобразования текстурированных солнечных элементов на основе кремния**

А.В. Саченко[1], В.П. Костылев[1], А.В. Бобыль[2], В.Н. Власюк[1], И.О. Соколовский[1], В.Н. Вербицкий[2], Е.И. Теруков[2,3], М.З. Шварц[2], М.А. Евстигнеев[4]

[1]*Институт физики полупроводников им. В.Е. Лашкарева НАН Украины, 03028 Киев, Украина, e-mail: sach@isp.kiev.ua,*

[2]*Физико-технический институт им. А.Ф. Иоффе, 194021 Санкт-Петербург, Россия*

[3]*НТЦ тонкопленочных технологий в энергетике, 194064 Санкт-Петербург, Россия*

[4] *Department of Physics and Physical Oceanography, Memorial University of Newfoundland, St. John's, NL, A1B 3X7 Canada*

Обнаружена трансформация механизмов формирования длинноволнового края внешнего квантового выхода в текстурированных солнечных элементах (СЭ) на основе кремния в зависимости от их толщины. Получены выражения, описывающие экспериментальные зависимости внешнего квантового выхода СЭ на длинноволновом краю поглощения в широком диапазоне толщин базы (100-450 мкм). Они позволяют рассчитать оптимальные значения толщины базы СЭ из условия максимума эффективности фотопреобразования при учете значений скорости поверхностной рекомбинации. В частности, установлено, что оптимальная толщина базы порядка 100 мкм достигается при скорости поверхностной рекомбинации около 3 см / с.

Повышение эффективности фотопреобразования $\eta$ является ключевой научно-практической задачей разработки солнечных элементов (СЭ) [1]. Увеличение $\eta$ достигается при уменьшении безызлучательной рекомбинации (объемной и поверхностной), а также при уменьшении коэффициента отражения света от фронтальной поверхности $r_s$. Уменьшение объемной рекомбинации достигается путем использования кремния с меньшей концентрацией дефектов, ответственных за рекомбинацию Шокли-Рида-Холла, и уменьшения толщины базы СЭ $d$. Уменьшение поверхностной рекомбинации происходит за счет создания на тыловой поверхности СЭ изотипного перехода или гетероперехода, а также благодаря уменьшению концентрации поверхностных рекомбинационных состояний из-за пассивирующего действия термического окисла кремния или внедренного в полупроводник водорода. Радикальным методом уменьшения $r_s$ является текстурирование одной или обеих поверхностей кремниевой пластины.

Важной задачей является оптимизация толщины СЭ. Величина $d_{opt}$ определяется из максимума произведения напряжения холостого хода $V_{OC}$ на плотность тока короткого замыкания $J_{SC}$. Можно показать, что для высокоэффективных СЭ справедливо выражение

$$V_{OC}(d) \propto \ln(J_{SC}(d)/q(d/\tau_b + S)), \quad (1)$$

где $q$ - элементарный заряд, $\tau_b$ - объемное время жизни, $S$ - суммарная скорость поверхностной рекомбинации на освещенной и тыльной поверхностях СЭ. Величина $J_{SC}(d)$ с увеличением $d$ растет, а зависимость $V_{OC}(d)$ определяется соотношением скоростей объемной и поверхностной рекомбинации, а также скоростью роста $J_{SC}(d)$.

Как показано в работе [2], вероятность поглощения фотона возрастает за счет увеличения его пути от значения $2d$ в плоскопараллельной кремниевой структуре с зеркальной тыловой поверхностью до значения $4n_r^2 d$ - в текстурированной, где $n_r$ - показатель преломления. Поэтому внутренний квантовый выход фототока во втором случае будет определяться не выражением для плоскопараллельной структуры вида

$$IQE_{nt}(\lambda) = 1 - \exp(-2\alpha(\lambda)d), \quad (2)$$

а формулой Яблоновича [2]

$$IQE_t(\lambda) = \left(1 + \left(4\alpha(\lambda)d\, n_r^2\right)^{-1}\right)^{-1}, \quad (3)$$

где $\alpha(\lambda)$ - коэффициент поглощения света в зависимости от длины световой волны $\lambda$. Выражения (2) и (3) справедливы только в случае высокоэффективных кремниевых СЭ, для которых $\eta \geq 20\%$. Они реализуются, когда имеют место следующие критерии: $L >> d$, $Sd/D << 1$, где $L$ и $D$ - длина и коэффициент диффузии неосновных носителей заряда.

На рис. 1а - 1в приведены экспериментальные зависимости $EQE(\lambda)$ для текстурированных $HIT$ элементов с толщиной, равной 98 мкм [3], измеренные в настоящей работе для $HIT$ элементов с толщиной 160 мкм и полученные в [4] для кремниевых СЭ с $p-n$ переходом при толщине, равной 450 мкм. На каждом из рисунков построены расчетные зависимости для внутреннего квантового выхода $IQE(\lambda)$, описываемые формулами (2) и (3). Величина коэффициента поглощения $\alpha(\lambda)$ взята из работы [5]. Для удобства расчетные кривые 1 и 2 нормированы на максимальное экспериментальное значение $EQE$. Значения толщины, фотоэнергетические параметры, объемные времена жизни и уровень легирования для использованных структур СЭ приведены в таблице 1.

Как видно из рис. 1а, только на нем есть участок, на котором зависимость $EQE(\lambda)$ описывается формулой Яблоновича (3). В общем случае экспериментальные результаты для зависимостей $EQE$, приведенные на рис. 1а – 1в, в области края поглощения могут быть согласованы с расчетом при использовании эмпирической формулы

$$IQE_s(\lambda) = \left(IQE_t x + IQE_{nt}(1-x)\right), (4)$$

где $x$ - численный коэффициент, изменяющийся между единицей и нулем. При $x \to 1$ имеет место формула Яблоновича (3), а в пределе $x \to 0$ справедливо выражение (2). Варьируя $x$, можно достичь удовлетворительного согласования расчетной зависимости (4) с экспериментом, приведенным на рис. 1а – 1в (см. кривые 3). Как видно из рис. 1, в текстурированных СЭ на основе кремния при увеличении толщины $d$ от значений ~100 мкм и выше происходит переход от механизма Яблоновича к механизму, связанному с однократным отражением света в плоскопараллельной структуре. Для структур СЭ с толщиной 98 мкм в области $\lambda \geq 1130$ нм эксперимент согласуется с формулой (3), а при $\lambda \leq 1000$ нм происходит постепенный переход к закону (2) (рис. 1а). Для структур СЭ с тол-щиной 160 мкм в области $1200 \geq \lambda \geq 1000$ нм работает комбинированный механизм, а при $\lambda \leq 1000$ нм происходит переход к механизму, справедливому для плоскопараллельных структур (рис. 1б). При толщинах ~ 450 мкм действует механизм, связанный с однократным отражением света в плоскопараллельной структуре (рис. 1.в). Заметим, что имеются и другие работы [6-8], в которых приведены зависимости $EQE(\lambda)$ СЭ на основе кремния, и в которых наблюдаются аналогичные результаты.

Величина $J_{SC}$ в условиях АМ1,5 определяется из выражения

$$J_{SC} = q \int_{\lambda_0}^{\lambda_m} I_{AM1,5}(\lambda) EQE(\lambda) d\lambda \ , (5)$$

где $\lambda_0$ - коротковолновый край поглощения, $\lambda_m$ = 1200 нм, $I_{AM1,5}(\lambda)$ - спектральная плотность потока фотонов при реализации условия АМ 1,5.

На рис. 2 приведены расчетные зависимости $J_{SC}(d)$, полученные с использованием (5). Кривые 1 - 3 получены с использованием экспериментальных значений для внешнего квантового выхода $EQE(\lambda)$, а также выражений (4) и (5). Для сравнения на рисунке построены зависимости $J_{SC}(d)$, полученные при использовании формулы (3) (см. кривые 1′ - 3′). Как видно из сравнения кривых (1-3) и (1′ - 3′), зависимости $J_{SC}$ при $d$ = 98 мкм близки к полученным из формулы Яблоновича, при $d$ = 160 мкм растут несколько быстрее, а при $d$ =450 мкм – заметно быстрее. Это согласуется с результатами, приведенными в [2].

Рассмотрим вопрос о соотношении вкладов в $J_{SC}$, связанных со сдвигом зависимости $EQE(\lambda)$ в длинноволновую сторону и с уменьшением коэффициента отражения $r_s$ в текстурированных СЭ. Используя выражение (5), и зависимости $EQE(\lambda)$ и $IQE(\lambda)$, приведенные на рисунках 1 и 2, можно оценить величину $\Delta J_{SC1}$, связанную с длинноволновым сдвигом. Относительная величина $\Delta J_{SC1}$ составляет около 7.5% для $d$ =98 мкм и около 2.8% для $d$ =160 мкм. Величина $\Delta J_{SC2}$, связанная с уменьшением $r_s$, может быть оценена при использовании формулы Френеля. Для кремния $r_s$ составляет около 30%, что приводит к росту $J_{SC2}$ примерно на 41% для случая [3], и на 38% при $d$ =160 мкм. При $d$ = 450 мкм [4], величина $J_{SC}$ текстурированных СЭ по сравнению с плоскими увеличивается на 41% исключительно за счет уменьшения $r_s$.

Для определения $\eta(d)$ воспользуемся результатами подхода к расчету параметров высокоэффективных СЭ на основе кремния, описанными в [9,10]. Данный подход позволяет рассчитать также величину $S$. Результаты расчета $S$ приведены в таблице 1. Используя выражения для $\eta$ из [9,10], данные для $S$ и рассчитывая зависимости $J_{SC}$ с использованием (4) и (5), получаем расчетные зависимости $\eta(d)$, приведенные на рис. 3. Значения $d_{opt}$ исследуемых СЭ приведены в таблице 1. Сравнивая толщины СЭ и величины $d_{opt}$ для разных СЭ, можно сделать вывод, что толщины исследуемых СЭ оптимизированы. Для достижения максимальных значений $\eta$ при толщинах ~100 мкм, необходимы низкие значения $S$, что характерно для $HIT$ СЭ. Как видно из таблицы 1, для исследованных $HIT$ элементов полученные значения $S$ равны соответственно 5,5 и 9 см/с. Для кремниевых СЭ с $p-n$ переходом значение $S$ равно 26 см/с. По-видимому, причиной этого является то, что пассивация поверхности с помощью водорода является более эффективной, чем пассивация с использованием термического окисла кремния.

Таким образом, в текстурированных высокоэффективных СЭ на основе кремния с увеличением их толщины происходит трансформация механизма формирования длинноволнового края $EQE(\lambda)$ от механизма формирования, описываемого формулой Яблоновича, к механизму, связанному с зеркальным отражением света в плоскопараллельной структуре. Значение оптимальной толщины, при которой реализуется максимальная величина $\eta$, критично к величине скорости поверхностной рекомбинации, в частности, $d_{opt}$ ~100 мкм может быть достигнуто, если $S \sim 3$ см/с.

Таблица 1

| Образец | d, мкм | $\tau_b$, мс | S, см/с | $J_{SC}$, мА/см² | $V_{OC}$, В | FF, % | $\eta$, % | $n_0$, см⁻³ | $d_{opt}$, мкм |
|---|---|---|---|---|---|---|---|---|---|
| № 1 [3] | 98 | 3 | 5,5 | 39,5 | 0,750 | 83,2 | 24,7 | $5 \cdot 10^{15}$ | 110-130 |
| № 2 Наши изм. | 160 | 1.5 | 9 | 36 | 0,721 | 78,6 | 20,4 | $1 \cdot 10^{15}$ | 190-210 |
| № 3[4] | 450 | 1.5 | 26 | 42 | 0,696 | 83,6 | 24,5 | $5 \cdot 10^{15}$ | 450-700 |

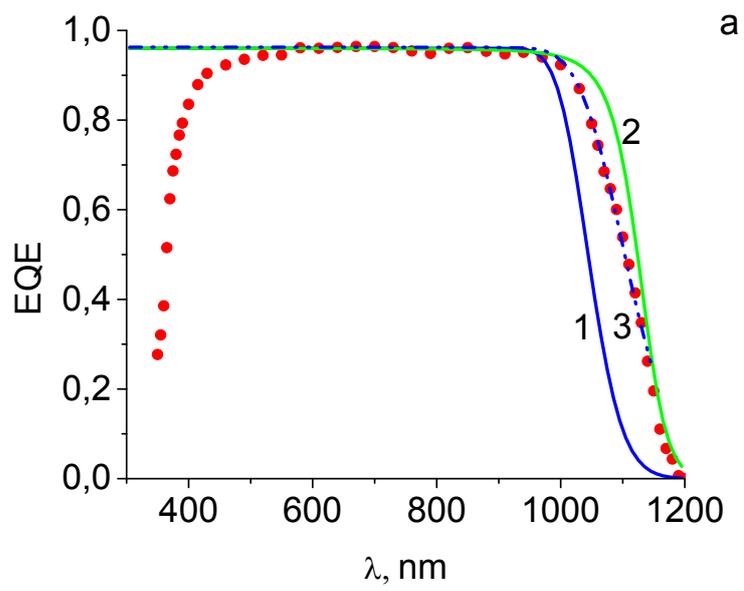

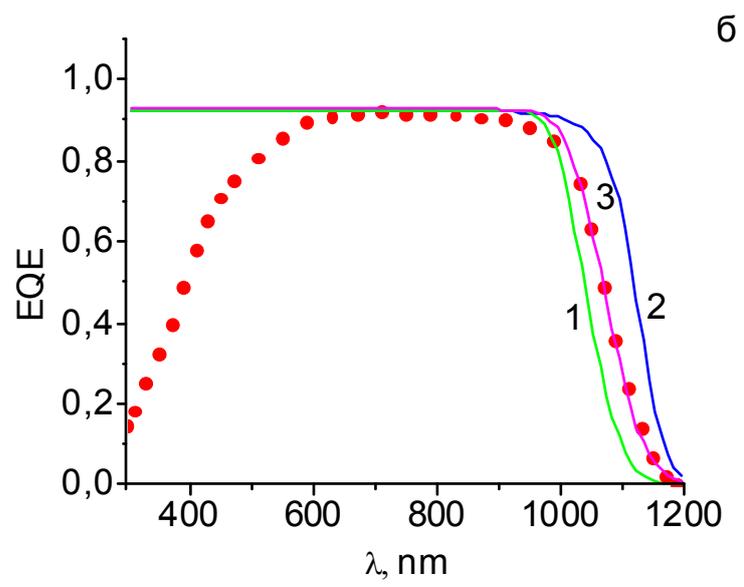

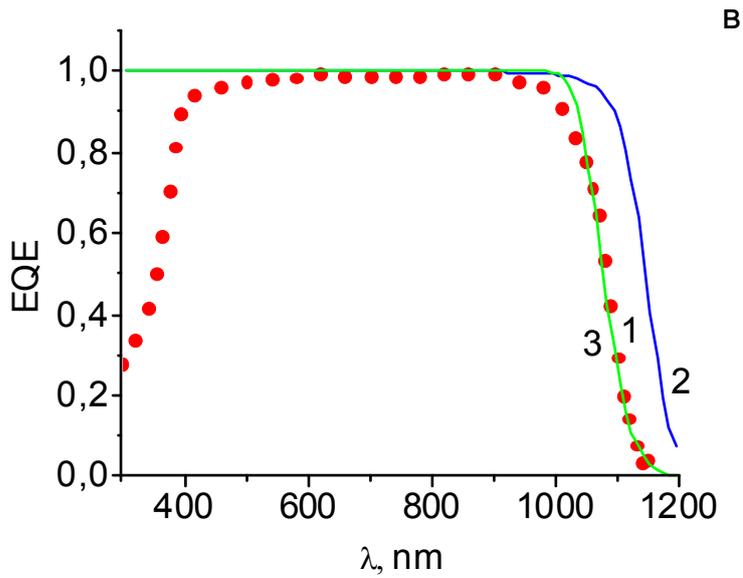

Рис. 1. Экспериментальные зависимости СЭ для внешнего квантового выхода, взятые из работы [3] (рис. 1 а), из наших измерений (рис. 1б) и из работы [4] (рис. 1в). Теоретические зависимости внутреннего квантового выхода (кривые 1-3) рассчитаны по формулам (2), (3) и (7). Использованные при расчете параметры: x=0.87 (рис. 1а), x=0.49 (рис. 1б) и x=0.03 (рис. 1в).

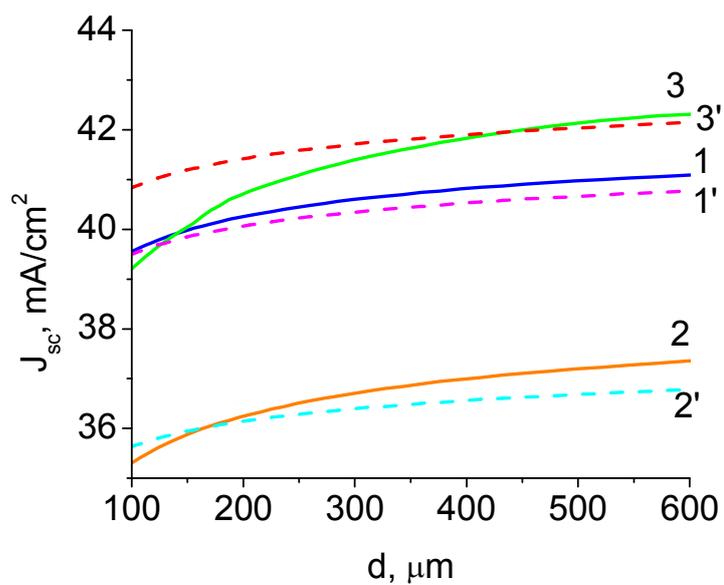

Рис. 2. Теоретические зависимости плотности тока короткого замыкания от толщины, построенные с использованием параметров, приведенных в таблице 1. Нумерация кривых в соответствии с таблицей 1. Штриховые кривые 1′ - 3′ построены с использованием формулы Яблоновича (3).

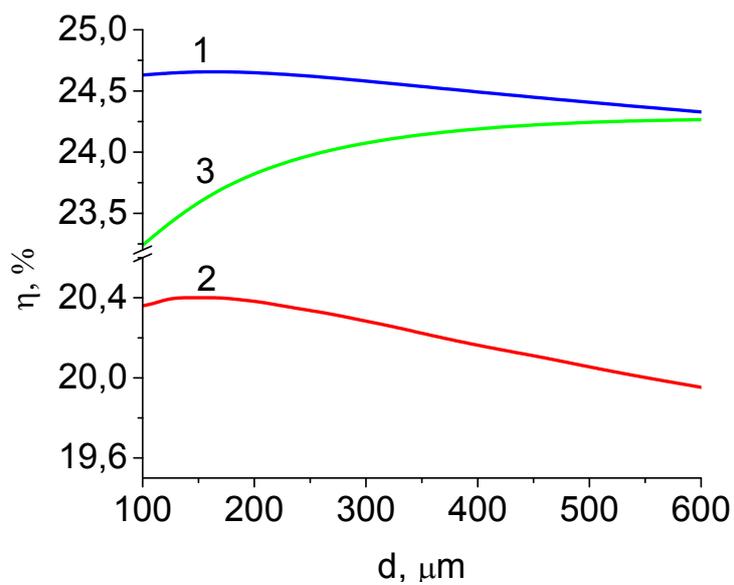

Рис. 3. Теоретические зависимости эффективности фотопреобразования от толщины, построенные с использованием параметров, приведенных в таблице 1. Нумерация кривых в соответствии с таблицей 1.